\documentclass[preprint2]{aastex}

\newcommand{\tc}{$T_{c}$}

\usepackage{color,natbib}

\shorttitle{TRAMOS: nine new transit epochs of the exoplanet WASP-5b}
\shortauthors{Hoyer et al.}

\begin{document}


\title{ Transit Monitoring in the South (TraMoS) project: Discarding Transit Timing Variations in WASP-5b}


\author{S. Hoyer\altaffilmark{1}}
\email{shoyer@das.uchile.cl}

\author{P. Rojo\altaffilmark{1}}
\email{pato@das.uchile.cl}

\and

\author{M. L\'opez-Morales\altaffilmark{2,3}}
\email{mercedes@dtm.ciw.edu}


\altaffiltext{1}{Astronomy Department, Universidad de Chile, Casilla
  36-D, Santiago de Chile, Chile}

\altaffiltext{2}{Institut de Ci\`{e}ncies de l'Espai (CSIC-IEEC),
  Campus UAB, Facultat de Ciencies, Torre C5, parall, 2a pl, E-08193
  Bellaterra, Barcelona, Spain}

\altaffiltext{3}{Visiting Scientist, Carnegie Institution of
  Washington, Department of Terrestrial Magnetism, 5241 Broad Branch
  Rd. NW, Washington, D.C. 20015, USA. }






\begin{abstract}

We report nine new transit epochs of the extrasolar planet
\object{WASP-5b}, observed in the \textit{Bessell I} band with SOAR at
the Cerro Pachon Observatory and with the SMARTS 1-m Telescope at
CTIO\footnote{Based on observations made with the SMARTS 1-m
    telescope at CTIO
    and the SOAR Telescope at Cerro Pachon Observatory under
    programmes ID CNTAC-08B-046,-09A-089 and  -09B-050.}
, between August 2008 and October 2009.  The new transits have
been combined with all previously published transit data for this
planet to provide a new Transit Timing Variations (TTVs) analysis of
its orbit.  We find no evidence of TTVs \textit{RMS} variations larger
than 1 min over a 3 year time
span.  This result discards the presence of planets
more massive than about $5 ~M_{\earth}$ ,  $1~M_{\earth}$  and  $2~M_{\earth}$  around the 1:2, 5:3 and
2:1 orbital resonances.  These new detection limits exceed by $\sim5-30$ times the limits 
imposed by current radial velocity observations in the Mean Motion
Resonances of this system. 
Our search for the variation of other parameters, such as orbital
inclination and transit depth also yields negative results over the
total time span of the transit observations. 
This result supports formation theories that predict a paucity of
planetary companions to Hot Jupiters. 

\end{abstract}


\section{Introduction}

Once the method of Transit Timing Variations (TTVs) was theoretical
proposed as of great potential to detect additional exoplanets in
transiting systems \citep{miralda02,agol05,holman05}, and even
exomoons \citep{sartoretti99,Kipping09a}, several observational groups
have started to monitor the majority of the known transiting planets.
This monitoring aims at detecting changes in the predicted mid-time of
the transits to infer the presence of additional planets in the system
not detected previously by e.g. radial velocities.  Those data have
also been used to detect variations in other transit parameters
(e.g. transit depth and duration), that can be attributed to
perturbations produced by unseen companions \citep{miralda02}.

In addition to having the potential of finding planets in the
Earth-like or smaller mass regime, the detection (or non-detection) of
companions of transiting Hot Jupiters through TTVs also can improve
constraints on  planet formation models \citep[e.g.][and references
  therein]{Triaud2010, Naoz2011, Miguel11} and help discriminate
between the different mechanisms proposed. In this way, TTVs become
a powerful tool for the detection and study of multi-planet
system architectures \citep{Latham2011,Ford.KeplerTTV.2011}.

The first unquestionable evidence of TTVs was announced by
\cite{Holman2010} in the double Saturn-like transiting planetary
system \object{Kepler-9}, where the central times of transit vary with
amplitudes of 4 and 39 minutes in timescales of about 19 and 40
days, respectively. Another extraordinary confirmation of the TTVs
effect came with the discovery of Kepler-11 \citep{Lissauer11}, a
system with six transiting planets which shows TTVs of amplitudes as
large as tens of minutes produced by the gravitational perturbations
between the planets.  An additional remarkable result of this later work has
been the use of dynamical orbital fits to determine directly the
masses of the transiting planets, dismissing the need of radial
velocities.

In 2008, the WASP-South survey reported their second detection of an
exoplanet, \object{WASP-5b}, transiting a relatively bright star
($V=12.3$) in the Southern Hemisphere
\citep[hereafter A08]{Anderson.WASP5.2008}. This discovery paper, based on WASP
photometry and two additional transit epochs plus radial velocities
measurements, announced a Hot-Jupiter planet with a mass of
$M_{P}=1.58^{+0.13}_{-0.08}~M_J$ and a density of
$\rho_{p}=1.22^{+0.19}_{-0.24} ~\rho_{J}$, orbiting a G4V star with a
period of $P=1.62$~days.

\cite{Gillon.WASP5.2009} did a reanalysis of the
A08 data to produce the first timing study of
\object{WASP-5b} and arrived to the conclusion of potential period
variations, based on a $\sim 2$-minute shift in the timing residuals
of the most precise points.
  
\cite{Southworth.WASP5.2009}, hereafter S09, observed two new transits, for which they
achieved very high photometric precision by defocusing the images at the
1.54-m Danish Telescope at La Silla Observatory, but
at the expense of producing only 3-minute cadence. They refined the
linear ephemeris of the system and concluded the high deviation of the
timing residuals with respect to that straight line
($\chi^{2}_{red}=5.7$) found by \cite{Gillon.WASP5.2009} was based on
the divergence of only one point out of six.
\cite{Smith.additExop.WASP.2009} searched for signatures of additional
planets in the residuals of WASP light curves after
removing the transits of \object{WASP-5b}, and found no evidence of a
transiting companion down to
Saturn-size planets within periods of up to 20 days.

Other recent works have determined and refined several physical
parameters of the system. For example, \cite{Triaud2010} determined
the angle between the orbital plane of \object{WASP-5b} and the
spin axis direction of its host star to be consistent with zero
($\lambda=12\degr^{+10}_{-8}$). This conclusion has been confirmed by
the reanalysis of \cite{Fukui11}, hereafter F11, who obtain $\lambda=7.2\degr\pm
9.5$.

F11 additionally searched for TTVs of
\object{WASP-5b} using seven new transit epochs, combined with all
previously available observations.  They find a $RMS$ of about 68 seconds in
their timing residuals despite of having an average of 41 seconds
uncertainty per epoch, and proposed that such a large deviation from
a linear fit ($\chi^{2}=32.2$ for 9 degrees of freedom) can be
explained by an orbital perturber.  Using dynamical simulations F11
constrained the masses of this hypothetical perturber to 
$2~M_{\earth}$ in the 1:2 and 2:1 mean motion resonances (MMRs) and
set a mass of $43~M_{\earth}$ for a potential Trojan body.  

\cite{Dragomir11}, hereafter D11, reported two new transits of WASP-5b with data of the 1-m telescope at Cerro Tololo Inter-American Observatory.  

In this work we present nine additional transits of \object{WASP-5b},
observed between August 2008 and October 2009, and perform a new
homogeneous timing analysis of all available epochs to further confirm
or rule out the TTV signals previously proposed for this system.

In section \ref{observaciones} we describe the new observations and
the data reduction. Section \ref{fiteo} details the modeling of the
light curves and in section \ref{ttv} we present the timing analysis.
In Section \ref{masslimits} we discuss the mass limits for a
\textit{unseen} perturber.  Finally, we present our conclusions in
section \ref{conclusiones}.

\section{Observations and data reduction \label{observaciones}} 

In 2008 we started the Transit Monitoring in the South Project,
which is a monitoring campaign of transiting planets observable from
the Southern Hemisphere \citep{Hoyer11}, following the approach of
using high-cadence observations and the same instruments and setups to
try to minimize systematics and reduce uncertainties in the
mid-transit times, as well as other transit parameters.  For the
TraMoS project we have already observed more than 60 transits of over
20 exoplanets.

As part of TraMoS we observed a total of nine transits of
\object{WASP-5b}, between August 2008 and October 2009\footnote{In the
  remaining of the text we refer to each individual transit by the UT
  date of mid-time of the transit, using the following notation
  YYYY-MM-DD}, with the Y4KCam on the SMARTS 1-m Telecope at Cerro
Tololo Inter-American Observatory (CTIO) and with the SOAR Optical
Imager (SOI) at the 4.2-meter Southern Astrophysical Research (SOAR)
telescope in Cerro Pach{\'on}.

Y4KCam is a $4064\times4064$ CCD camera with a Field of View (FoV) of
$20\times20$ squared arcminutes and a pixel scale of 0.289 arcsec
pixel$^{-1}$.  The standard readout time of the camera is $46$~sec,
which we reduce to $\sim16$~sec by binning 2x2. The SOI detector is
composed of two E2V mosaics of $4096\times4096$ pixels with a scale of
0.077 arcsec pixel$^{-1}$, giving a FoV of $5.2\times5.2$ squared
arcminutes.  The instrument has a $20.6$~sec standard readout, which
becomes only $\sim11$~ sec after binning 2x2.

\begin{figure*}
\epsscale{1.75} 
\plotone{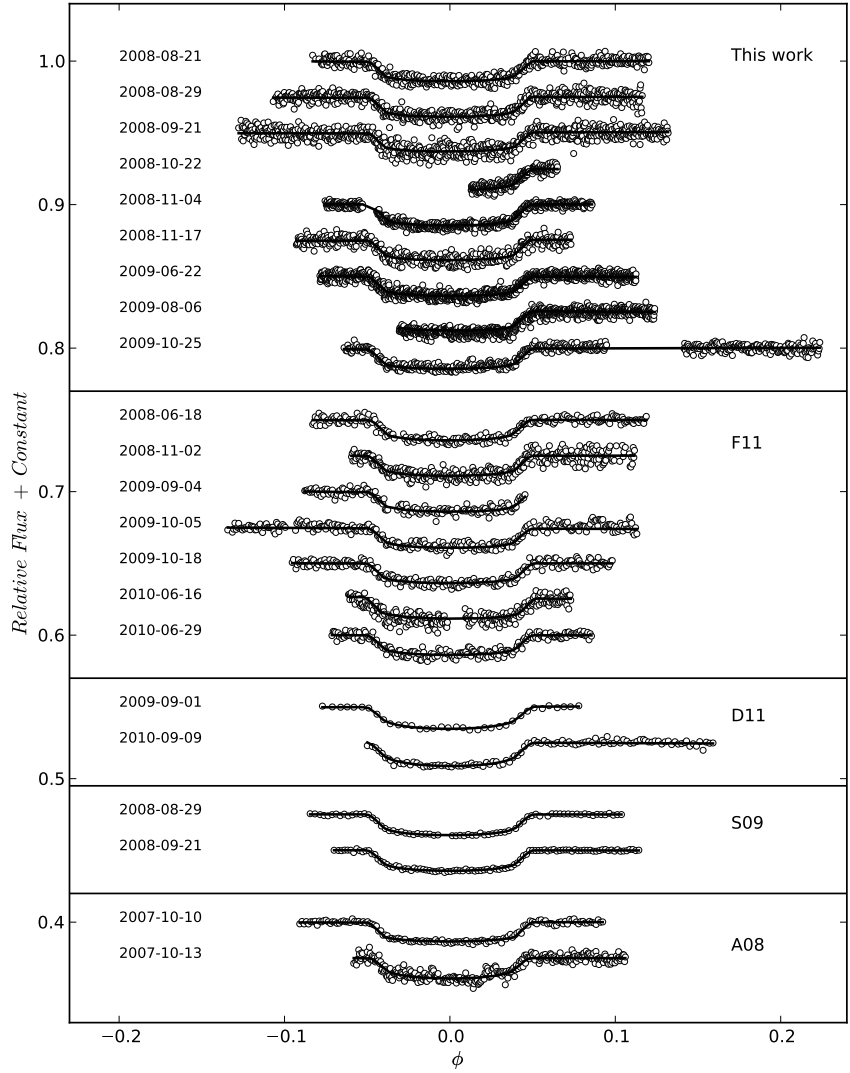}
\caption{Light curves of the nine transits of WASP-5b presented in
  this work, the seven transits of F11, the two transits of D11, S09 and A08. The solid lines
  show our best model fits using TAP (see section
  \ref{fiteofinal}). The UT date is indicated in the left of each
  light curve. \label{lcall}}
\end{figure*}

All nine transits were observed using a \textit{Bessell I} filter
($\lambda_{\rm eff}=8665~\AA$ and FWHM=$3914~\AA$) to reduce limb
darkening effects in our light curves.  Six of the transits were fully
covered in phase. A fraction of the ingress of the 2008-11-03 transit
was not observed because a telescope system crash as illustrated in
Figure \ref{lcall}.  Nevertheless, this
transit was treated as a \textit{complete} transit.  Two other
transits, 2009-08-05 and 2009-10-21, were only partially covered with
data between phases $-0.034 ~\lesssim \phi \lesssim ~0.01$ and $0.12
~\lesssim \phi \lesssim ~0.06$, respectively.   Two of our transits,
2008-08-29 and 2008-09-21, coincide with the transit epochs published by
S09.  The observing log is summarized in
Table~\ref{tabla-obs}.

The initial trimming, bias and flatfield corrections of all the
collected data were performed using custom-made pipelines specifically
developed for each instrument.  The times at the start of the exposure
are recorded in the image headers, in particular we used the value of
the \textit{Modified Julian Day} (JD-2400000.5) field.  In
the SMARTS telescope, the time stamp recorded in the header of each frame is
generated by a \textit{IRIG-B GPS time synchronization protocol}
connected to the computers that control the instrument.  The SOAR
telescope data use
the time values provided by a time service connected to the
instrument.  We confirmed that these values have $\sim1$~second
precision. The time value assigned to each frame corresponds to the
\textit{Julian Day} at the start of the exposure plus $1/2$ of the
integration time of each image (see section \ref{ttv} for details).

\object{WASP-5} is located in a relative empty field, where both the
target and several well suited comparison stars appear well isolated
in our images.  Therefore, we extracted the flux from the target and
comparison stars via standard aperture photometry, and using our own
python-based code. We used a range of stellar apertures between 8 and
12 pixels, and sky rings which extended between 25 and 35 pixels in
radius.

For each sky-aperture combination, we generated differential light
curves between the target and each comparison star to 1) optimize the apertures and 2)
select the best comparison stars.  The criterium used in both cases
was \textit{RMS} minimization for the out-of-transit and in-transit data (excluding the ingress and egress portions of the light curves).  The final light curves were generated
computing the ratio between target's flux and the best 2 to 5
comparison stars.

Finally, some systematics remaining after this step were removed by means of
linear or quadratic regression fits to the out-of-transit light curve
points using X-Y pixel position, time and/or airmass as free parameters. The final light
curves present average photometric dispersions of the order of $0.2 \%$ - $ 0.45\%
$.

\section{Light Curve Modeling \label{fiteo}}

\subsection{Algorithm Comparison\label{comparacion}}

 We performed a comparison between algorithms that use
different statistical uncertainty estimation techniques to the transit's
parameters, in order to test
potential systematics between them. 
  There are different approaches to do that statistical 
error estimation analysis; for example,
JKTEBOP\footnote{http://www.astro.keele.ac.uk/~jkt/codes/jktebop.html}
\citep{South04a,South04b} uses the
Levenberg-Marquardt Monte Carlo (LMMC) technique to compute errors  \citep[see e.g.][]{Southworth-LCIII-2010, Hoyer11},
while several other studies have started to implement Monte Carlo Markov Chains
(MCMC) techniques \citep[e.g.][]{Adams2010, Fulton11}.

In \cite{Hoyer11} we proposed that the results of both, the LMMC and
MCMC algorithms are equivalent if the parameter space lacks of local
minima, where LMMC minimization can be trapped.  Here we further test
that proposal by comparing the results of both algorithms on the
\object{WASP-5b} data used for this study.  We compare the results of
fitting a light curve of WASP-5b with JKTEBOP and the Transit Analysis
Package\footnote{http://ifa.hawaii.edu/users/zgazak/ifA/TAP.html}
\citep[TAP;][]{TAP}, which implements the MCMC method for the estimation
of errors \citep[more details in][and references therein]{Fulton11}.

Among the parameters that JKTEBOP allows to fit are: the
planet-to-star radius ratio ($R_p/R_s$), the inclination ($i$) and
eccentricity ($e$) of the orbit, the out-of-transit baseline flux
($F_{oot}$), the mid-time of transit ($T_c$), the quadratic limb
darkening coefficients ($\mu_1$ and $\mu_2$), and the sum of the
fractional radii,  $R=R_p / a + R_s /
a$, where $R_p$ and $R_s$ are the absolute stellar and planetary radii, and
$a$ is the orbital semi-major axis. TAP allows to fit for all those
parameters except for the latter, which is replaced by $a/R_s$.

For the comparison we left free all the mentioned parameters except
$a/Rs$ and $R$ in TAP and JKTEBOP, respectively, since they
otherwise presented convergence problems.  We also fixed $F_{OOT}=1$, $e=0$ and
$\mu_2(I)=0$ and the orbital period to $P=1.62843142$~days from F11
since any variation in this parameter will be detected later in our
timing analysis.   We used $10^4$ iterations in
JKTEBOP and 10 chains of $10^5$ steps each in TAP.  We discarded the
first $10\%$ iterations on each chain to compute the final parameter's
values and its respective errors.   The results on each fit, shown in Table \ref{tablaTAPvsJKTEBOP},
reveal that the resultant fit values of all parameters common to the
JKTEBOP and TAP algorithms agree within the error, except for
$\mu_1(I)$.  

\begin{figure*}
\plotone{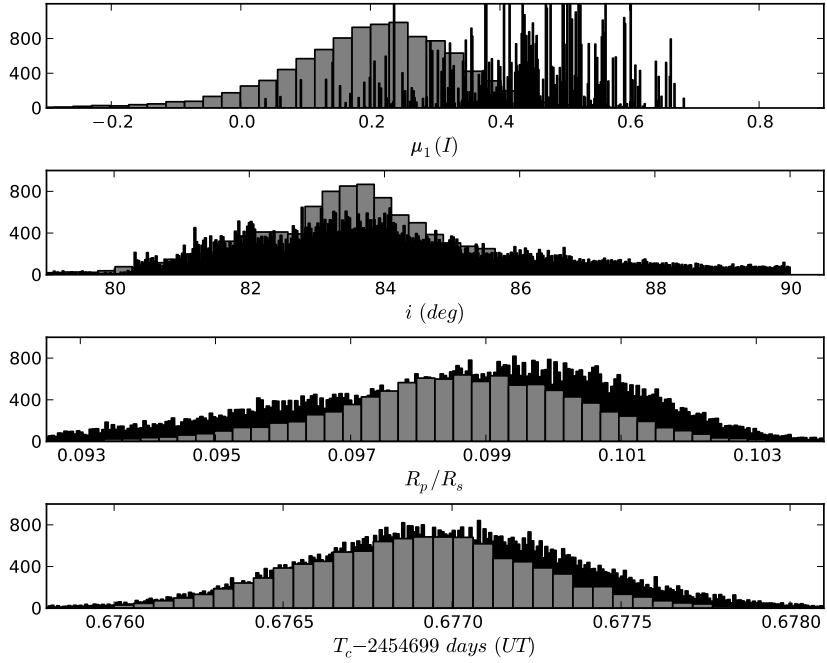}
\caption{Histograms of the 10\,000 LMMC iterations with JKTEBOP (gray
  histograms) and of the 10 chains with 10\,000 links (after
  discarding the first $10\%$ results on each chain) obtained with TAP (black
  histograms) for the 2008-08-21 epoch.  
For comparison, the binning factor of the TAP results histograms is 9
times the binning factor used for the JKTEBOP results. In both fittings,
all the parameters were left free except for quadratic limb darkenning coefficients ($\mu_2(I)=0$), eccentricity ($e=0$) and the periastron longitude ($\omega=0$). \label{comp-distribuciones}}
\end{figure*}

\begin{figure*}
\plotone{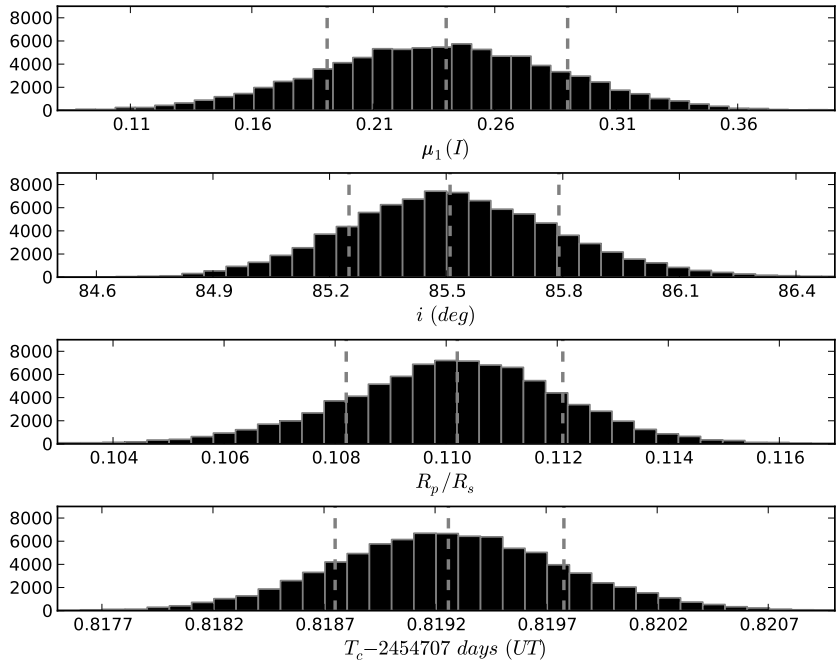}
\caption{Histograms of the Markov Chain Monte Carlo iterations
  resulting of fitting the 2008-08-29 transit data with TAP, for their
  fitted parameters: $\mu_{1}(I)$, $R_{p}/R_{s}$, $i$ and \tc. The dashed lines show the
  fitted value and the $\pm 1\sigma$ errors. \label{histograms}}

\end{figure*}

Figure \ref{comp-distribuciones} shows the
distribution of each parameter obtained using the LMMC and the MCMC
techniques with data from 2008-08-21 transit (similar analysis
was done with the other 6 complete light curves).  From the three bottom
panels in the Figure \ref{comp-distribuciones} it is evident that the
$1\sigma$ errors (defined as the 68\% of a Gaussian fit to the parameter value
distributions) obtained with LMMC are generally smaller than those
obtained using MCMC,
since the latter  does a more exhaustive exploration of the parameter space and
therefore  performs better error estimations.  Also, from the top-panel
of Figure  \ref{comp-distribuciones}, it can be seen that the
LMMC results for  certain parameters  can appear biased towards
their initial input values. That is the case for the linear
limb-darkening coefficient, for which the value resulting from the
LMMC analysis is $\mu_1(I)=0.22\pm0.12$ (the initial value was
0.296).  On the other hand, the
distribution of values for  this parameter on a single epoch as given by MCMC does not
appear Gaussian, revealing that the quality of a single transit in the
current data does not allow to constrain the values of $\mu_1(I)$.
Notice, however, that a Gaussian distribution is obtained when fitting
several transits simultaneously (see Figure \ref{histograms} and section
\ref{fiteofinal}).

From the test results above we conclude that the LMMC and MCMC 
techniques arrive to similar parameter results.  However, because the
apparent underestimation of the errors estimated by LMMC we have opted
for using TAP for our analysis of the full \object{WASP-5b} transit
dataset and the re-analysis of all the available data (see next
section).    This underestimation is due to lack of multi-parameter uncertainty estimator and failure 
to account for red noise in the minimization \citep{Carter09} as TAP does.    Other advantages of TAP include that the code can fit a greater number of parameters like linear systematics in
the datasets, and it allows a simultaneous
fitting of multiple transits.  

\subsection{Final Modeling \label{fiteofinal}}

We used TAP to fit the nine new transit light curves presented in this
paper and all the available light curves of the system (seven of F11, two of D11, two of S09
and the two of A08). 

First, we attempted to model each of the new light curves
independently, but ran into several problems. TAP had difficulties
fitting the incomplete light curves. Also, when fitting individual
light curves, parameters such as $\mu_1$ did not clearly converge to a
single value, as already mentioned in section \ref{comparacion} and
illustrated in Figure \ref{comp-distribuciones}. To avoid these
problems we fit the seven complete light curves simultaneously,
leaving as free parameters $\mu_1(I)$, $i$, $R_p/R_s$, $T_c$,
$F_{OOT}$, in addition to possible linear trends to the light curves,
$F_{slope}$, and white (uncorrelated) and red (correlated) noise
components, $\sigma_w$ and $\sigma_r$, respectively. The orbital
period, the eccentricity, and the longitude of the periastron were
fixed to the values P= 1.62843142 days (the value obtained by F11),
$e=0$ and $\omega = 0$.

We used a quadratic limb-darkening law, but found that the precision
of the light curves was not enough to reliably fit the quadratic
coefficient, so that value was also fixed to $\mu_2(I) = 0.32$, based
on the tabulated results in \citet{Claret2000}.

As mentioned in Section \ref{observaciones}, we initially corrected for systematic trends
in the light curves using linear or quadratic regression fits.  Altough slopes in the light curves
are not clearly apparent, we leave $F_{OOT}$ and $F_{slope}$ as free parameters to ensure that any
small residuals are properly fit.   This might create concerns about wheter this two-step fitting of 
systematics can affect the results of the fits.   To ensure we are not introducing any bias on the 
determination of the planetary parameters, we fit the two sets of data (i.e. the light curves with and without systematics
trends removed) with TAP and arrive to consistent values of all derived planetary parameters.

\begin{figure*}
\plotone{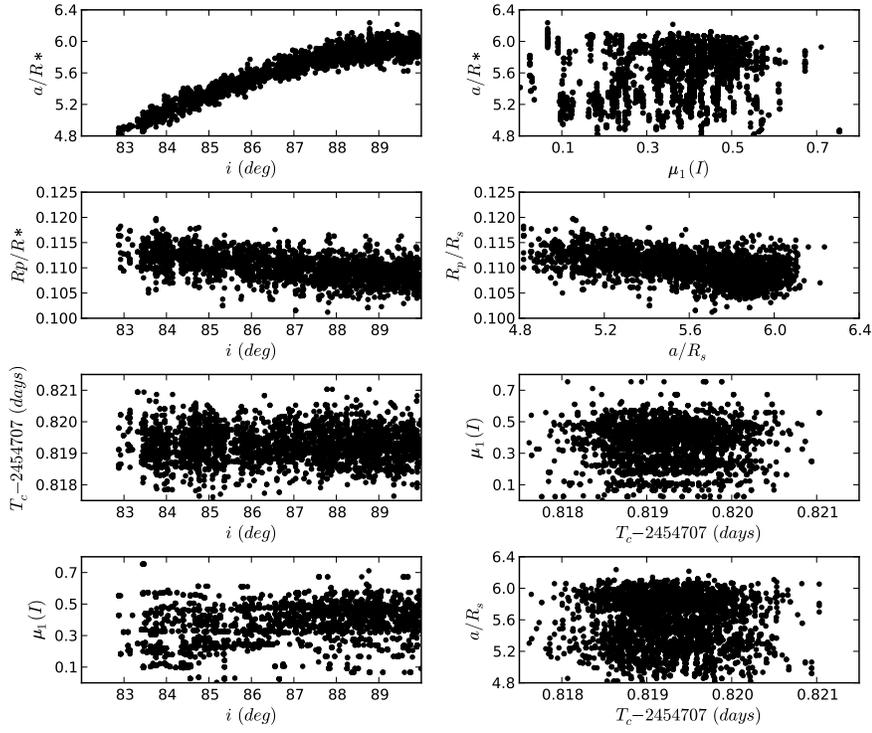}
\caption{Results of the Markov Chain Monte Carlo
  iterations resulting of fitting the 2008-08-29 transit data with
  TAP, which show the correlation  between the fitted light curve parameters $a/R_s$,
  $R_{p}/R_{s}$, \tc, $i$ and 
  $\mu_{1}(I)$. \label{distribuciones}}
\end{figure*}

We also searched for potential parameter correlations in the light
curves using the fit results of the 2008-08-29 transit described in
the previous section, where all the parameters were let to vary. The
resultant parameter correlations are shown in Figure
\ref{distribuciones}. This figure reveals a strong correlation between
$a/R_s$ and $i$. There is also evidence of weaker correlations between
those two parameters and $R_p/R_s$. Therefore, to minimize the impact
of those correlations in our results, we fixed $a/R_s$ in all the
light curves to 5.37 (from F11), while closely monitoring $R_p/R_s$
and $i$ for variations.

To fit the transits we ran 10 MCMC chains of $10^5$ links each,
discarding the first 10\% results from each chain to avoid bias
toward the initial input values of each fitted parameter. Because the
resulting MCMC distributions for $\mu_1(I)$ are not Gaussian (see
Figures \ref{comp-distribuciones} and \ref{distribuciones}), that
parameter was fit simultaneously for all seven light curves, while for
the other parameters we obtained one value per curve and combine them
afterward via a weighted average. The resulting average values for
each parameter are listed in Table \ref{Tabla-parametros}, together
with their 1$\sigma$ errors. As an example, Figure \ref{histograms}
shows the resultant MCMC distributions of $i$, $R_p/R_s$, and $T_c$
for the transit observed on 2008-08-29, while the
distribution of $\mu_1(I)$ correspond to the results of the
simultaneous seven transits fit. This distribution is now clearly
Gaussian in contrast with the previously obtained.

Finally, we adopted the values of all the parameters that define the
shape of the transit derived in the fit above and used them as fixed
values in the two incomplete light curves (2008-10-22 and 2009-08-06
transits) to derive their mid-times of transit, $T_c$. The $F_{OOT}$,
$F_{slope}$, $\sigma_w$ and $\sigma_r$ are still left variable in this case.

F11 used a procedure based on $\chi^2$ minimization for modeling their
light curves.  We re-analyzed their data to do an homogeneous study of
all the light curves, given that a multi-parameter minimization based
on MCMC is statistically more robust. We modeled the seven light
curves of F11\footnote{The data is available in the on-line material
  from the F11 publication on PASJ}, the two light curves of D11 (data provided by the author, private communication), the two light curves of S09\footnote{The data is
  available at the CDS (http://cdsweb.u-strasbg.fr/)} and the two of
A08 (data provided by the author, private communication) in a similar manner to our
complete light curves above. The F11 transits were  observed with
a \textit{Bessel I} filter, the D11 and the S09 with a $R$ filter,  and
the A08 with $R$ and SDSS \textit{i'} filter; therefore, we fit one $\mu_1
(I)$ simultaneously for all F11 curves, one $\mu_1(R)$ for the D11
curves and one for the S09 curves, and separate $\mu_1(i)$ and $\mu_1(R)$ for the A08 curves. We fixed $\mu_2=0.32$ in all cases.


The obtained parameters are summarized Table \ref{Tabla-parametros}.
The resultant models to all 22 light curves are illustrated in Figure \ref{lcall}.

We point out that the errors of the F11's light curves estimated by us are, in average, $70~\%$ larger than the reported by F11.   We checked that the origin of this difference was not due only by the different red-noise estimator methods.
Using the same red-noise factor estimated by F11, we have obtained errors consistent with those we present in Table \ref{Tabla-parametros}.   \cite{Carter09} found that time averaging and residual permutation methods underestimated the errors by $~15-30\%$ compared with the wavelet-based method (implemented by TAP).

\begin{figure*}
\plotone{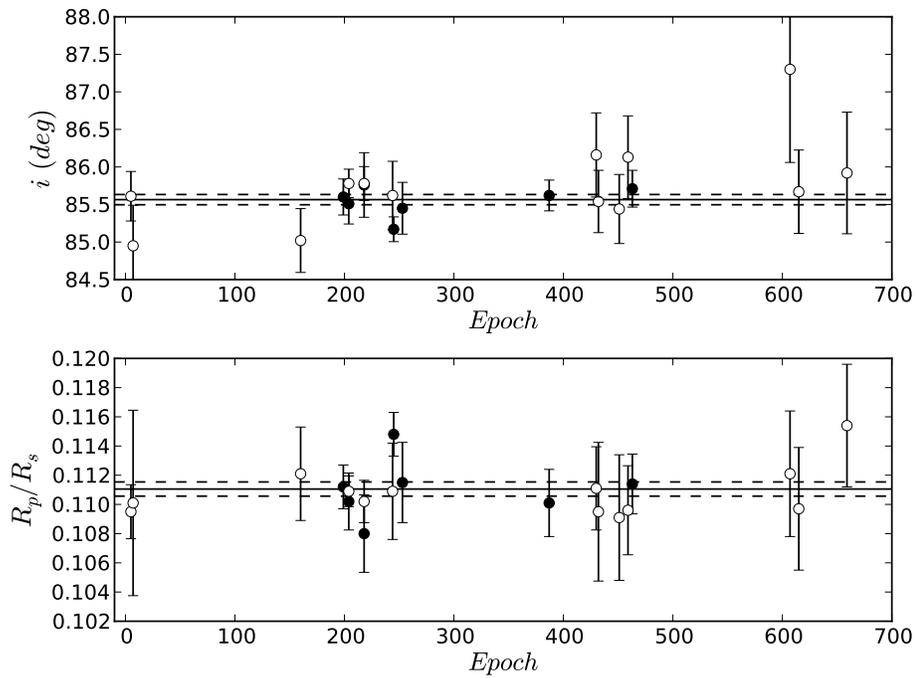}
\caption{Derived values of the orbital inclination (\textbf{Top
    Panel}) and planet-to-star radii ratio, $R_p/R_s$ (\textbf{Bottom
    Panel}) for all modeled transits. The solid circles
  correspond to our seven complete transits. The open circles
  correspond to the seven transits of F11 and the two
  transits of D11, S09 and A08.  The weighted
  average to all points is represented by the solid line on each panel
  and the dashed lines show the $\pm 1\sigma$ errors of those
  fits. No significant variations are apparent for these parameters
  in the time span of the observations.  \label{parametros} }

\end{figure*}

Using the model results is
possible to look for variations in the most relevant parameters, in
particular $i$ and $R_p/R_s$, that can reveal the presence of an
additional body in the system.  In Figure \ref{parametros}, we plot
$R_p/R_s$ and $i$ as a function of the transit epoch, based in the
results of the twenty transit fits (our two incomplete light curves were not included). We do not see any significant
variations in those parameters. The weighted average values of $i$ and
$R_p/R_s$ based on all the light curves results are summarized in
Table \ref{tabla-final}. We studied in detail the timing of the
transits in the next section.

\section{Timing Analysis \label{ttv}}

The times in our nine transit data and the D11 data were initially computed in
Coordinated Universal Time (UTC) and then converted to Barycentric
Julian Days, expressed in Terrestrial Time, BJD(TT), using the
\cite{Eastman2010} online
calculator \footnote{http://astroutils.astronomy.ohio-state.edu/time/utc2bjd.html}. The
transit times of S09 and A08, which were initially expressed in HJD(UT)
have also been converted to BJD(TT). No conversion was
applied to the light curves reported by F11.   

The times of the common transits, 2008-08-29 and 2008-09-21, derived from our
light curves are consistent within the errors in the values derived by
us and also by F11 from S09 data.  

\begin{figure*}
\plotone{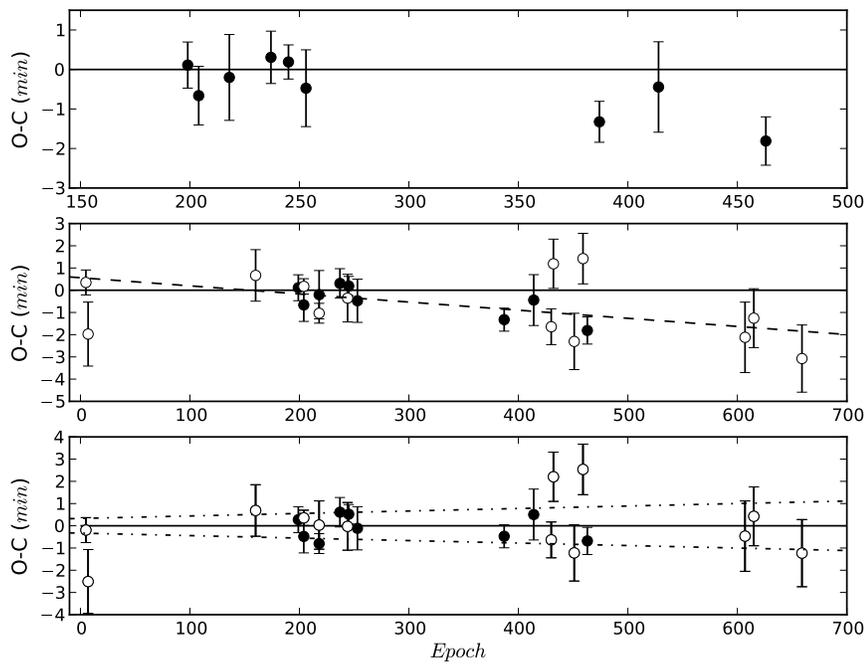}
\caption{\textbf{Top-panel:} \textit{Observed minus Calculated} diagram
  of the central times of the nine transits reported in this work.
  \textbf{Middle-panel:} $O-C$ residuals of our nine transits (solid
  circles) combined with the $O-C$ residuals of the new fits to the
  F11, D11, S09 and A08 transits (open circles).  The dashed line shows the linear trend found in the data.
  \textbf{Bottom-panel:} $O-C$ residuals of all available data after
  removing the linear trend.  The solid and point-dashed lines in this
  figure correspond, respectively, to our new ephemeris equation fit
  and its associated $\pm 1 \sigma$ errors.  \label{o-c}}
\end{figure*}

Using the F11's ephemeris equation, we calculated the
residuals of the mid-times of the 22 transits of \object{WASP-5b}
analyzed in this work.
The top panel in Figure \ref{o-c}, shows the $Observed$ $minus$
$Calculated$ ($O-C$) diagram for our nine transits. In the middle
panel of the figure we combine the $O-C$ values of our nine transits
with the new values derived for the F11, D11, S09 and A08 (shown as open
circles).  As illustrated in that
figure, a linear trend with a slope of $2.54 \times 10^{-6} $~days is observed in the time
residuals of all transits.  That trend can be explained by the
accumulation of errors in the current orbital period and $T_{0}$ of
the transits over time, and therefore can be modeled out.

This linear regression of the points in the $O-C$ diagram has a 
$\chi^2_{red}=1.22$ ($\chi^2=24.37$ for 20 degrees of freedom), which
is significantly smaller than the value obtained for F11 of $\chi^2_{red}=3.66$
($\chi^2=32.2$ for 9 degrees of freedom).  Additionally, we confirmed that with our results for the 11 epochs included in F11's analysis we also obtained an smaller $\chi^2$ ($\chi^2=15.45$ that yields $\chi^2_{red}=1.72$).  This result lies in the fact that our $T_{c}$ uncertainties are larger than those estimated by F11.      

Once the linear trend is removed the updated ephemeris equation is:

\begin{eqnarray*}
T_{c} = 2454375.62459(23)[BJD_{TT}] + \nonumber \\
         1.62842888(78)  \times E,  
\end{eqnarray*}

where \tc ~is the central time of a transit in the epoch $E$ since the
reference time $T_{0}$. The errors of the last digits are shown in
parenthesis.  
The bottom panel in Figure \ref{o-c} shows the resulting $O-C$ values
of all available transits using the updated ephemeris equation. The
resultant $O-C$ diagram is consistent with a constant period, and we
conclude that the observed TTV residuals (with a \textit{RMS} of 
$\sim 0.00073$~days  $\simeq 63$  seconds), are most likely introduced by
data uncertainties and systematics rather than
due by gravitational perturbations of an orbital companion.  
This newly obtained precision  permits to place strong constraints in the mass of an
hypothetical companion, particularly in MMR's, as we discuss in the
next section.

\section{Limits to additional planets \label{masslimits}}

To place upper limits to the potential perturbers in
the \object{WASP-5} system based in the derived TTV \textit{RMS} of
about 60 sec we use \textit{Mercury} \citep{Chambers99} N-body
simulator. The input parameters to $Mercury$ include the mass and the radius of
both the star and the transiting planets, the planet-to-star orbital
separation, as well as the inclination, eccentricity and periastron
longitude of the system.  The values for all these
parameters were adopted from S09. In
addition, all the initial relative angles between
the perturber and \object{WASP-5b} were set to zero.  

We explored a wide range of perturber masses between $1~M_{\earth}$ and
$4000~M_{\earth}$ in initial steps of $50~M_{\earth}$, which are
subsequently refined as described below. For the semi-major axis
distances we explore a range between 0.001 and  1.2 AU in steps of
0.001 AU, which was further reduce near resonances.  The density of
the perturber was kept constant to  that of Earth for $M_p \leq 10
M_{\earth}$ and to that of Jupiter for $M_P \geq 200 M_{\earth}$, it was
varied linearly for masses in between. Also, we assumed the perturber to be in a
circular orbit and coplanar to \object{WASP-5b}, since this configuration
provides the most strict limit to the amplitude of the TTVs for a
given perturber's mass. Non-zero eccentricities and non-coplanar
orbits produce larger TTVs as already pointed out by
e.g. \cite{Bean09}, \cite{Hoyer11} and \cite{Fukui11}. For each model configuration we
let the system relax for five years, and then we used the next five
years to obtain our fit results, which in total is
more than 3 times the time span of the observations.  These 5 years of
relaxation time permits to minimize the effect of any initial bias (e.g. the
relative angles).  We found orbits between 0.02 and 0.035 ~AU to be
unstable due to the presence of WASP-5b.  For all other (stable)
orbits we recorded the central times of each transit of WASP-5b and
computed the predicted TTVs for each configuration, assuming an average
constant period. Additionally, we checked that the fitted average period did not
deviate by more than $3~\sigma$ from the obtained orbital period of
\object{WASP-5b}.  Also, to ensure a good sampling of the potential
perturber's mass, we reduced the steps in $M_{pert}$ to $1
M_{\earth}$ whenever the TTVs approached 60 sec.  

\begin{figure*}
\plotone{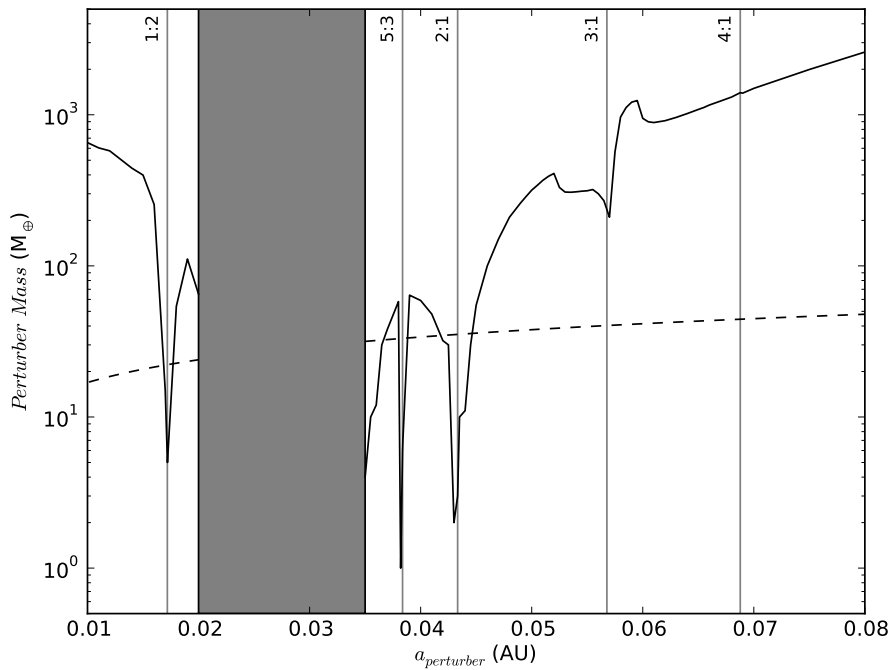}
\caption{Upper mass limits of an orbital perturber derived by
  dynamical simulations done with \textit{Mercury} code \citep{Chambers99}.  The solid
  line represents transit timing variations limits with a \textit{RMS} of 1 minute.
  The dashed line corresponds to the limits imposed by the current radial velocities
  observations. Vertical lines indicate the location of the MMR
  distances with WASP-5b for orbital separations of less than 0.08
  AU.  The gray band indicates the range of distances in which any other
  object would be in a unstable orbit.  \label{mvsa}}
\end{figure*}

The results of our model simulations is illustrated in Figure
\ref{mvsa}, where we show the $M_{pert}$ ($M_{\earth}$) versus $a
(AU)$ diagram that places the mass limits to potential perturbers in
the system. The solid line in the diagram indicates the derived upper
limits to the mass of the perturbers that would produce TTVs \textit{RMS} of
$60$~sec at different orbital separation.  The dashed line shows the
perturber mass upper limits imposed by the most recent radial
velocity observations of the WASP-5 system, for which we have adopted
a precision of 15 m/s \citep[A08 and][ report RV precision of 14~m/s and
$12-18$~m/s, respectively]{Triaud2010}.

Figure \ref{mvsa} thus shows that the perturber would have been detected by RV
measurements in all areas except around the 1:2, 5:3 and 2:1  MMRs,
where it could have a maximum mass of 5, 1 and 2  ~$M_{\earth}$, respectively.

\section{Conclusions \label{conclusiones}}

We present nine new transit light curves of \object{WASP-5b}.  We
homogeneously model these light curves together with all available
transit data of this system. Based in these fits we
search for any variation in the timing of the transits. 

Using 22 transit epochs we updated the ephemeris equation and we
find a TTVs \textit{RMS} of $63$ ~seconds.  All the transit times are
consistent with a constant orbital period within 2$\sigma$.

Our linear fit of the transit times has a $\chi^2_{reduce}=1.22$,
which is considerably lower than the value found by \cite{Fukui11} used to
implied the presence of an perturber body.  

Despite obtaining a similar TTV \textit{RMS} than \cite{Fukui11} ($\sim1$ min),
we conclude a much smaller significance to deviations from a constant
period due to our larger  per-epoch uncertainties as obtained by the MCMC
algorithm. 

If the system has an additional orbiting body, its mass has to be lower than
5, 1 and 2 $M_{\earth}$, in the 2:1, 5:3 and
1:2 resonances. In any other location the perturber would have been
detected by RVs.

We search for any trend in the depth of the transit and inclination of
the orbit but we do not see any clear evidence of variation with
statistical significance.

\section{Acknowledgements}

The authors would like thank David Anderson for providing WASP light curves and
Diana Dragomir and Stephen Kane for providing TERMS light curves.  
S.H. and P.R. acknowledgements support from Basal PFB06, Fondap
\#15010003, and Fondecyt \#11080271. Additionally, S.H, recieved
support from ALMA-CONICYT FUND \#31090030.  
We thanks the CTIO and SOAR staff for the help and continuous support during the
numerous observing nights.

\bibliographystyle{apj}
\bibliography{ttv-refs,utils-refs}

\clearpage

\begin{deluxetable}{lclclc}
\tabletypesize{\scriptsize}
\tablecaption{Observational information of each night.\label{tabla-obs}}
\tablewidth{0pt}
\tablehead{
\colhead{Transit Date} & 
\colhead{Telescope/Instrument} &
\colhead{Filter} &
\colhead{Integration Time [s]} &
\colhead{airmass range} &
\colhead{Epoch}
}
\startdata
2008-08-21 &  SMARTS-1m/Y4KCam          & Bessell I &  13           & 1.7 - 1.01         & 199\\
2008-08-29\tablenotemark{a} & SMARTS-1m/Y4KCam  & Bessell I &   10          & 1.05 - 1.02 - 1.06 & 204   \\
2008-09-21\tablenotemark{a} &  SMARTS-1m/Y4KCam & Bessell I&   10,7 & 1.9 - 1.01 - 1.07   & 218   \\
2008-10-22\tablenotemark{b} & SOAR/SOI   &  Bessell I &   7,5,3      & 1.12 - 1.02        & 237 \\
2008-11-04 &   SOAR/SOI                  & Bessell I &   3           & 1.07 - 1.02 - 1.4  & 245 \\
2008-11-17 &   SMARTS-1m/Y4KCam         & Bessell I &   10          & 1.02 - 1.4         & 253 \\
2009-06-22 & SOAR/SOI                    & Bessell I &   7,5,3       & 1.95 - 1.02        & 387 \\
2009-08-06\tablenotemark{b} & SOAR/SOI   & Bessell I &   5,4         & 1.07 - 1.15        & 414 \\
2009-10-25 & SMARTS-1m/Y4KCam           & Bessell I &   15          & 1.06 - 1.02 - 1.97 & 463 \\
\enddata
\tablenotetext{a}{This transit was also observed by \cite{Southworth.WASP5.2009}.}
\tablenotetext{b}{This transit has a incomplete phase coverage.}

\end{deluxetable}

\clearpage

\begin{deluxetable}{lcccccc}

\tablewidth{0pt}
\tablecaption{Values obtained with Levenberg-Marquardt Monte Carlo (JKTEBOP) and Markov Chain
  Monte Carlo (TAP) algorithms with data of the 2008-08-21 transit of WASP-5b.\label{tablaTAPvsJKTEBOP}}
\tablehead{
\colhead{Parameter} &
\colhead{JKTEBOP} & 
\colhead{TAP} 
}

\startdata

$R_{p}/R_{s}$ &  $0.0988 \pm 0.0018 $   &  $0.0988 \pm 0.0026$\\
$i ~[\degr]$ &  $83.4 \pm 1.5$         &  $83.7 \pm 2.3 $ \\
$\mu_{1}(I)$     &  $0.22 \pm 0.12$      & $0.45 \pm 0.11$\\
$T_{c}-2454699$ ($UT$) & $0.67690 \pm .00035 $  & $0.67697 \pm 0.00041 $\\
$(R_p+R_s)/a$ & $0.223 \pm 0.015 $ & \nodata \\
$ a /R_s $ & \nodata & $5.01 \pm 0.48$\\

\enddata

\end{deluxetable}

\clearpage

\begin{deluxetable}{lcccccc}
\tabletypesize{\scriptsize}
\tablewidth{0pt}
\tablecaption{Adjusted parameters for each transit using TAP.\label{Tabla-parametros}}
\tablehead{
\colhead{Transit date} &
\colhead{Epoch} &
\colhead{$R_{p}/R_{s}$} & 
\colhead{$i ~[\degr]$} & 
\colhead{$\mu_{1}(X)$ \tablenotemark{a} }& 
\colhead{$T_{c}-2450000$ ($BJD_{TT}$)} &
\colhead{$\sigma_{red}$/$\sigma_{white}$}
}

\startdata

2008-08-21  &  199 &$0.1112_{-0.0015}^{+0.0015}$   & $85.60^{+0.25}_{-0.23}$
& $0.237^{+0.05}_{-0.049} $ & $4699.68303^{+0.00040}_{-0.00041}$ &
1.6 \\ 

2008-08-29  & 204 &$0.1102_{-0.0020}^{+0.0019}$   & $85.51^{+0.28}_{-0.26}$
& $0.237^{+0.05}_{-0.049} $ & $4707.82465^{+0.00052}_{-0.00051}$ & 
2.8 \\ 

2008-09-21  & 218 &$0.1080_{-0.0026}^{+0.0027}$   & $85.76^{+0.46}_{-0.40}$
& $0.237^{+0.05}_{-0.049} $ & $4730.62301^{+0.00075}_{-0.00076}$ &
2.9 \\ 

2008-10-22  & 237 &$0.1116$\tablenotemark{b}    & $85.47$\tablenotemark{b}
& $0.24 $\tablenotemark{b} & $4761.56356^{+0.00047}_{-0.00045}$ &
2.3 \\ 

2008-11-04  & 245 &$0.1148_{-0.0015}^{+0.0015}$   & $85.17^{+0.17}_{-0.16}$
& $0.237^{+0.05}_{-0.049} $ & $4774.59093^{+0.00030}_{-0.00030}$ &
5.3 \\

2008-11-17  & 253 &$0.1115_{-0.0028}^{+0.0027}$   & $85.45^{+0.36}_{-0.33}$
& $0.237^{+0.05}_{-0.049} $ & $4787.61792^{+0.00069}_{-0.00066}$ &
2.5 \\ 

2009-06-22  & 387 &$0.1101^{+0.0022}_{-0.0024}$   & $85.62^{+0.21}_{-0.20}$
& $0.237^{+0.05}_{-0.049} $ & $5005.82714^{+0.00036}_{-0.00036}$ &
10.3 \\

2009-08-06  & 414 &$0.1116$\tablenotemark{b}   & $85.47$ \tablenotemark{b}&
$0.24 $\tablenotemark{b} & $5049.79540^{+0.00080}_{-0.00079}$ &
8.2 \\ 

2009-10-25  & 463&$0.1114^{+0.0020}_{-0.0021}$   & $85.71^{+0.26}_{-0.23}$
& $0.237^{+0.05}_{-0.049} $ & $5129.58759^{+0.00042}_{-0.00043}$ &
5.6 \\ 

&&&&&&\\
\tableline
&&&&&&\\


2008-06-18\tablenotemark{c}& 160  & $0.1121^{+0.0032}_{-0.0032}$   &
$85.02^{+0.44}_{-0.41}$ & $0.292^{+0.089}_{-0.089} $ &
$4636.17459^{+0.00079}_{-0.00082}$ &  2.9 \\ 

2008-11-02\tablenotemark{c}& 244  & $0.1109^{+0.0034}_{-0.0032}$   &
$85.62^{+0.50}_{-0.41}$ & $0.292^{+0.089}_{-0.089} $ &
$4772.96212^{+0.00074}_{-0.00075}$ &  2.2 \\ 

2009-09-04\tablenotemark{c}& 432  & $0.1095^{+0.0048}_{-0.0047}$   &
$85.54^{+0.45}_{-0.38}$ & $0.292^{+0.089}_{-0.089} $ &
$5079.10830^{+0.00075}_{-0.00079}$ &  2.0 \\ 

2009-10-05\tablenotemark{c}& 451  & $0.1091^{+0.0041}_{-0.0045}$   &
$85.44^{+0.50}_{-0.42}$ & $0.292^{+0.089}_{-0.089} $ & $5110.04607
^{+0.00087}_{-0.00089}$ & 10.1 \\ 

2009-10-18\tablenotemark{c} & 459 & $0.1096^{+0.0030}_{-0.0031}$   &
$86.13^{+0.63}_{-0.47}$ & $0.292^{+0.089}_{-0.089} $ & $5123.07611
^{+0.00079}_{-0.00079}$ & 2.5 \\ 

2010-06-16\tablenotemark{c} &607  & $0.1121^{+0.0044}_{-0.0042}$   &
$87.30^{+1.5 }_{-0.98}$ & $0.292^{+0.089}_{-0.089} $ & $5364.0815
^{+0.0011 }_{-0.0011 }$ & 4.9 \\ 

2010-06-29\tablenotemark{c}&615   & $0.1097^{+0.0040}_{-0.0044}$   &
$85.67^{+0.63}_{-0.48}$ & $0.292^{+0.089}_{-0.089} $ & $5377.10955
^{+0.00091}_{-0.00093}$ & 5.2 \\ 

&&&&&&\\
\tableline
&&&&&&\\


2009-09-01\tablenotemark{d}&430  & $0.1111^{+0.0028}_{-0.0029}$   &
$86.16^{+0.59}_{-0.53}$ & $0.51^{+0.11}_{-0.13} $ & $5075.84947
^{+0.00056}_{-0.00056}$ & 8.0 \\

2010-09-09\tablenotemark{d}&659   & $0.1154^{+0.0041}_{-0.0043}$   &
$85.92^{+0.94}_{-0.68}$ & $0.51^{+0.11}_{-0.13} $ & $5448.75927
^{+0.0010}_{-0.0011}$ & 4.6 \\

&&&&&&\\
\tableline
&&&&&&\\


2008-08-29\tablenotemark{e}&204   & $0.1109^{+0.0011}_{-0.0010}$   &
$85.78^{+0.20}_{-0.18}$ & $0.367^{+0.052}_{-0.053} $ & $4707.82523
^{+0.00023}_{-0.00025}$ & 3.0 \\

2008-09-21\tablenotemark{e}&218   & $0.1102^{+0.0014}_{-0.0015}$   &
$85.78^{+0.24}_{-0.25}$ & $0.367^{+0.052}_{-0.053} $ & $4730.62243
^{+0.00031}_{-0.00031}$ & 3.7 \\

&&&&&&\\
\tableline
&&&&&&\\


2007-10-10\tablenotemark{f}& 5   & $0.1095^{+0.0017}_{-0.0020}$   &
$85.61^{+0.37}_{-0.29}$ & $0.37^{+0.11}_{-0.1} $ & $4383.76750
^{+0.00038}_{-0.00040}$ & 4.9 \\

2007-10-13\tablenotemark{f}& 7   & $0.1101^{+0.0061}_{-0.0066}$   &
$84.95^{+0.59}_{-0.49}$ & $0.39^{+0.18}_{-0.21} $ & $4387.0.2275
^{+0.0010}_{-0.0010}$ & 11.7 \\

\\

\enddata

\tablenotetext{a}{In all the fits the quadratic coefficient was
  fixed to  $\mu_{2}=0.32$.}
\tablenotetext{b}{These parameters were fixed in the modeling and
  correspond to the weighted average  of the results of the other
  seven full phase covered light curves presented in this work.}
\tablenotetext{c,d,e,f}{Fitting results of the transits of
  \cite{Fukui11} , \cite{Dragomir11}, \cite{Southworth.WASP5.2009} and
  \cite{Anderson.WASP5.2008}, respectively. }
%
\end{deluxetable}

\clearpage

\begin{deluxetable}{lcc}
\tablecaption{ Improved orbital values derived from the weighted
  average of the light curve's fits. \label{tabla-final}}
\tablewidth{0pt}
\tablehead{
\colhead{Parameter} & 
\colhead{Adopted Value} &
\colhead{ $1\sigma$~Error} 

}
                    \startdata
$ a/R_{S}$ \tablenotemark{a}     & $5.37$ & $\pm 0.15 $ \\
$R_{p}/R_{s}$   & $0.1111 $    &  $\pm 0.0005 $      \\
$i ~[\degr]$  &  $85.56 $    &  $\pm 0.07 $      \\
Period [days] &  $1.62842888$   &  $\pm 0.00000078  $        \\
$T_{o} ~[BJD_{TT}]$ & $2454375.62549 $  &  $\pm 0.00023  $       \\

\enddata

\tablenotetext{a}{Value and error adopted from \cite{Fukui11}.}

\end{deluxetable}

\end{document}